\documentclass{article}

\usepackage{authblk} 
\usepackage{amsmath,amssymb,graphicx,mathrsfs,xcolor,babel,textcomp,esint}

\title{Coulomb potential determining terahertz polarization in two-color laser field}

\author[1,5]{Yuan Gao}
\author[3,6]{Yizhu Zhang}
\author[4]{Kaixuan Zhang}
\author[1,5]{Ziyang Gan}
\author[1,7]{Tian-Min Yan}
\author[1,2,5,*]{Yuhai Jiang}

\affil[1]{Shanghai Advanced Research Institute, Chinese Academy of Sciences, Shanghai 201210, China}
\affil[2]{Center for Transformative Science and School of Physical Science and Technology, ShanghaiTech University, Shanghai 201210, China}
\affil[3]{Center for Terahertz Waves and College of Precision Instrument and Optoelectronics Engineering, Key Laboratory of Opto-electronics Information and Technical Science, Ministry of Education, Tianjin University, China}
\affil[4]{Zhejiang Provincial Key Laboratory of Ultra-Weak Magnetic-Field Space and Applied Technology, Hangzhou Innovation Institute, Beihang University, Hangzhou, 310051, China}
\affil[5]{University of Chinese Academy of Sciences, Beijing 100049, China}

\affil[6]{e-mail: zhangyizhu@tju.edu.cn}
\affil[7]{e-mail: yantm@sari.ac.cn}
\affil[*]{e-mail: jiangyh3@shanghaitech.edu.cn}

\begin{document}
\maketitle

\begin{abstract}
The orientation and ellipticity of terahertz (THz) polarization generated by two-color strong field not only cast light on underlying mechanisms of laser-matter interaction, but also play an important role for various applications. We develop the Coulomb-corrected classical trajectory Monte Carlo (CTMC) method to well reproduce the joint measurements, that the THz polarization generated by the linearly-polarized 800 nm and circularly-polarized 400 nm fields is independent on two-color phase delay. The trajectory analysis shows that the Coulomb potential twists the THz polarization by deflecting the orientation of asymptotic momentum of electron trajectories. Further, the CTMC calculations predict that, the two-color mid-infrared field can effectively accelerate the electron rapidly away from the parent core to relieve the disturbance of Coulomb potential, and simultaneously create large transverse acceleration of trajectories, leading to the circularly-polarized THz radiation. 

\end{abstract}

\section{Introduction}

The manipulation of the orientation and ellipticity of the THz electric field has broad applications in imaging the macromolecular chiral structures of proteins and DNA \cite{niessen_protein_2019, peng_three-dimensional_2020} and spectroscopy \cite{tielrooij_structure_2009, matsuda_room-temperature_2020}. An intense and broadband THz pulse can be generated by focusing two-color strong laser fields in gas medium, and the THz polarization can be directly manipulated by controlling the laser-gas interaction process \cite{cook_intense_2000}. Because the polarization devices are avoided in this process, the THz polarization can be engineered without the limitation of bandwidth and intensity threshold. 

In two-color scheme, the THz polarization can be experimentally manipulated by controlling the relative orientation \cite{zhang2020, yu_fedorov_thz_2017} and phase delay \cite{2009Coherent1, PhysRevLett.103.119903} between the fundamental ($\omega$) and second-harmonic (2$\omega$) electric fields, and the THz ellipticity can also be experimentally engineered by controlling the length of plasma filament \cite{2018Manipulation}. Meanwhile, the theoretical investigations provide several new strategies to manipulate the THz polarization and ellipticity \cite{bogatskaya_circularly_2020, PhysRevLett.115.049902}. However, all the above methods rely on regulating the laser electric fields to tune the THz polarization. In the strong-field induced THz radiation, besides the laser electric fields, the intrinsic Coulomb field of gas medium is simultaneously presented in the process. The open question is that, whether the Coulomb field influences the THz polarization characteristics? The issue is essentially relevant to the application in optical attoclock by means of THz radiation \cite{babushkin_2022}.

The Coulomb field extensively manifests itself in strong-field phenomena, including Coulomb focusing effect \cite{brabec_coulomb_1996}, 
atomic holographic imaging \cite{huismans_time-resolved_2011} and low-energy structure  \cite{2009Strong,PhysRevLett.103.119901}. Analogously, the THz generation in two-color scheme, which is attributed to Brunel radiation \cite{brunel_harmonic_1990}, 
is a typical strong-field phenomenon as well, where Coulomb effect should be taken into account. So far, the Coulomb-corrected phase-delay dependence of THz yields has been experimentally observed and delicately investigated \cite{2012Synchronizing,L2013Attosecond,PhysRevA.92.033410,PhysRevA.102.023109}.

In this paper, based on the trajectory-based Coulomb-strong field approximation \cite{Yan2010,Huang2019}, the Coulomb correction is incorporated into the trajectory method for THz radiation \cite{Fan:23,zhang_electron_2021}, which makes CTMC compatible with THz radiation. We found that the Coulomb potential is able to twist the THz polarization, which manifests itself under the specific polarization combination of the two-color laser fields, and the Coulomb effect is experimentally verified in our measurement. The trajectory analysis is used to explain the twist of THz polarization by means of the morphology of electron trajectory. Finally, the CTMC method is applied to predict the polarization behaviors of THz generation in mid-infrared two-color fields, which can be approximately incidated by the guiding center of electron trajectory \cite{Dubois2018}.

\section{Experimental procedure}

\begin{figure}[h]
\centering
    \includegraphics[width=0.8\linewidth]{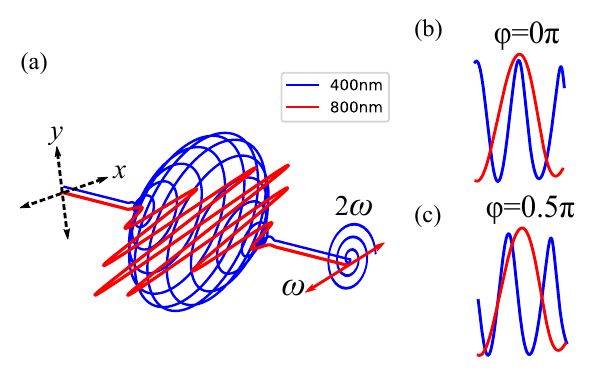}
\caption{Schematics of two-color electric fields. (a)  Circularly-polarized $\omega$ laser field and linearly-polarized 2$\omega$ laser field. The orientation of $\omega$ electric field is aligned along $x$-axis. (b) and (c) Two-color laser fields when the phase delay is \begin{math} \varphi = 0\pi \end{math} (b) and \begin{math} \varphi = 0.5\pi \end{math} (c).}
\label{fig: Two-color laser field}
\end{figure}
In the experiment, a Ti:sapphire femtosecond laser delivers 800 nm, 35 fs, 1.2 mJ laser pulses. The 800 nm ($\omega$) beam passes through a type-I \begin{math} \beta\end{math}-barium borate (\begin{math} \beta\end{math}-BBO) crystal to generate 400 nm (2$\omega$) beam with the efficiency of $\sim$30$\%$. The $\omega$-2$\omega$ beams are separated by a actively-stabilized Michelson interferometry, and propagate along the two arms of the interferometry, which is able to individually control the polarization of $\omega$ and 2$\omega$ beams. During the measurement, the $\omega$ electric field is longitudinally aligned along the $x$-axis, and the 2$\omega$ electric field is circularly polarized. By moving the BBO along the propagation axis, the relative phase delay $\tau$ between the $\omega$-2$\omega$ pulses is tunable with the accuracy of sub-femtosecond. The $\omega$-2$\omega$ beams are focused by a parabolic mirror with focus length of 100 mm to produce the plasma, where the THz wave emits from. The THz electric fields along both orthogonal polarizations, $E_{\mathrm{THz}}^{\mathrm{x}}$ and $E_{\mathrm{THz}}^{\mathrm{(y)}}$, are obtained by the electro-optical sampling detection. The time trace of the THz instantaneous electric field can be reconstructed with $E_{\mathrm{THz}}^{\mathrm{(x)}}$ and $E_{\mathrm{THz}}^{\mathrm{(y)}}$ \cite{zhang_portraying_2009}. The details of the experimental setup refer to \emph{Supplementary Material}.

\section{Theoretical methods}

According to the experimental geometry, in the CTMC simulation, the circularly 2$\omega$ field and linearly polarized $\omega$ field are combined. The $\omega$ electric field is aligned along the $x$-axis, and the laser
propagation direction is along the $z$-axis. The two-color laser fields can be expressed as
\begin {equation}\begin{aligned} \boldsymbol{E} (t)&= [A_{{\omega}}\cos ({\omega}t)+A_{2{\omega}}\cos (2{\omega}t+\varphi)]f(t)\boldsymbol{{y}}\\ 
&+A_{2{\omega}}\sin (2{\omega}t+\varphi)f(t)\boldsymbol{{x}}, \end{aligned}\end {equation}
where \begin{math} A_{{\omega}}\end{math} and \begin{math} A_{2{\omega}} \end{math} are the amplitudes, and \begin{math} A_{{\omega}} : A_{2{\omega}}=\sqrt3:1 \end{math}. \begin{math} f(t) =\sin^2(\frac{{\omega}t}{2n})\end{math} is the pulse
envelope, where $n$ is the cycle number of $\omega$ pulse. Here, the 800 nm pulse consists of 15 cycles, and the 400 nm pulse has the same duration of 30 cycles. And \begin{math} \varphi \end{math} is the relative phase between $\omega$-2$\omega$ fields.

We assume that the electron tunnels out along the instantaneous polarized axis at ionization instant \begin{math} t_{{0}}\end{math}. The ionization rate is given by
\begin{equation} W_{0}(t_{0})=C_l^2\frac{2l+1}{2}(\frac{2\kappa^2}{\left|{E(t_0)}\right|})^{\frac{2}{\kappa}-1} \text{exp}[\frac{-2\kappa^3}{\left|{E(t_0)}\right|}], \end{equation}
where \begin{math} {E(t_0)} \end{math} is the instantaneous electric field amplitude, and
\begin{math} \kappa=\sqrt{2I_\text{p}}\end{math} with \begin{math} I_\text{p}\end{math} being the atomic ionization potential. The
normalization factor \begin{math} C_l\end{math} is assumed as proposed by Tong $et$ $al$ 
\cite{PhysRevA.66.033402}.
In addition to the ionization rate, the initial momentum distribution after tunneling ionization needs to be considered. The initial longitudinal momentum (along the instantaneous laser polarization) is zero and the initial transverse momenta \begin{math} p_\bot\end{math} (perpendicular to the laser
polarization) are distributed with the probability according to
Ammosov, Delone and Krainov (ADK) tunneling theory \cite{1986Tunnel}:
\begin {equation} W_{1}(p_\bot)=\frac{p_\bot}{\pi}\frac{\kappa}{\left|{E(t_0)}\right|}\text{exp}[\frac{-\kappa p^2_\bot}{\left|{E(t_0)}\right|}]. \end {equation}
Therefore, the tunneling rate should be assigned to each classical trajectory. That is, \begin{math} W(t_0,p_\bot)=W_{0}(t_{0})W_{1}(p_\bot).\end{math}
For the convenience of simulation, we rewrite the initial momentum frame as $p_{\text{x}}=-p_\bot\cos{\alpha}\sin{\theta}$, $p_{\text{y}}=p_\bot\cos{\alpha}\cos{\theta}$, $p_{\text{z}}=p_\bot\sin{\alpha}$. Here \begin{math} \alpha \end{math} is the angle of \begin{math} p_\bot \end{math} from the positive $z$ axis, while \begin{math} \theta \end{math} is the angle between the polarized axis and the $x$-axis.
After tunneling, the electrons move under the action of laser and Coulomb fields, and the Newtonian equation of the electron reads,
\begin {equation} \boldsymbol{\Ddot{r}}=-\frac{\boldsymbol{r}}{r^3}-\boldsymbol{E} (t), \end {equation}
where $\boldsymbol{r}$ is defined as the electron trajectory, and $r$ is the scalar of $\boldsymbol{r}$. The initial position of the tunneling electron will be $x_{0}=r_{0}\cos{\theta}, y_{0}=r_{0}\sin{\theta}, z_{0}=0$. Here, $r_{0}=-\frac{I_\text{p}}{E(t_{0})}$ is the tunneling exit of the electron.
The radiation electric field is the temporally superposition of light waves induced by electron trajectories \cite{Fan:23,zhang_electron_2021}, written as
\begin {equation}\boldsymbol{E}_{\text{rad}}(t)\propto \sum\limits_\eta^{} {{W^{(\eta )} (t_\theta,p_\bot)} \boldsymbol{\Ddot{r}}}^{(\eta )}, \end {equation}
where \begin{math}{W^{(\eta )} (t_0,p_\bot)}\end{math} describes the ionization probability of the $\eta$th electron trajectory. The THz radiation \begin{math}\boldsymbol{E}_{\text{THz}}\end{math} is acquired by filtering out the Fourier components \begin{math} \mathcal{F}\left\{\boldsymbol{E}_{\text{rad}}(t)\right\} \end{math}  below 20 THz.

\section{Discussion}

\begin{figure}[h]
\centering
    \includegraphics[width=0.8\linewidth]{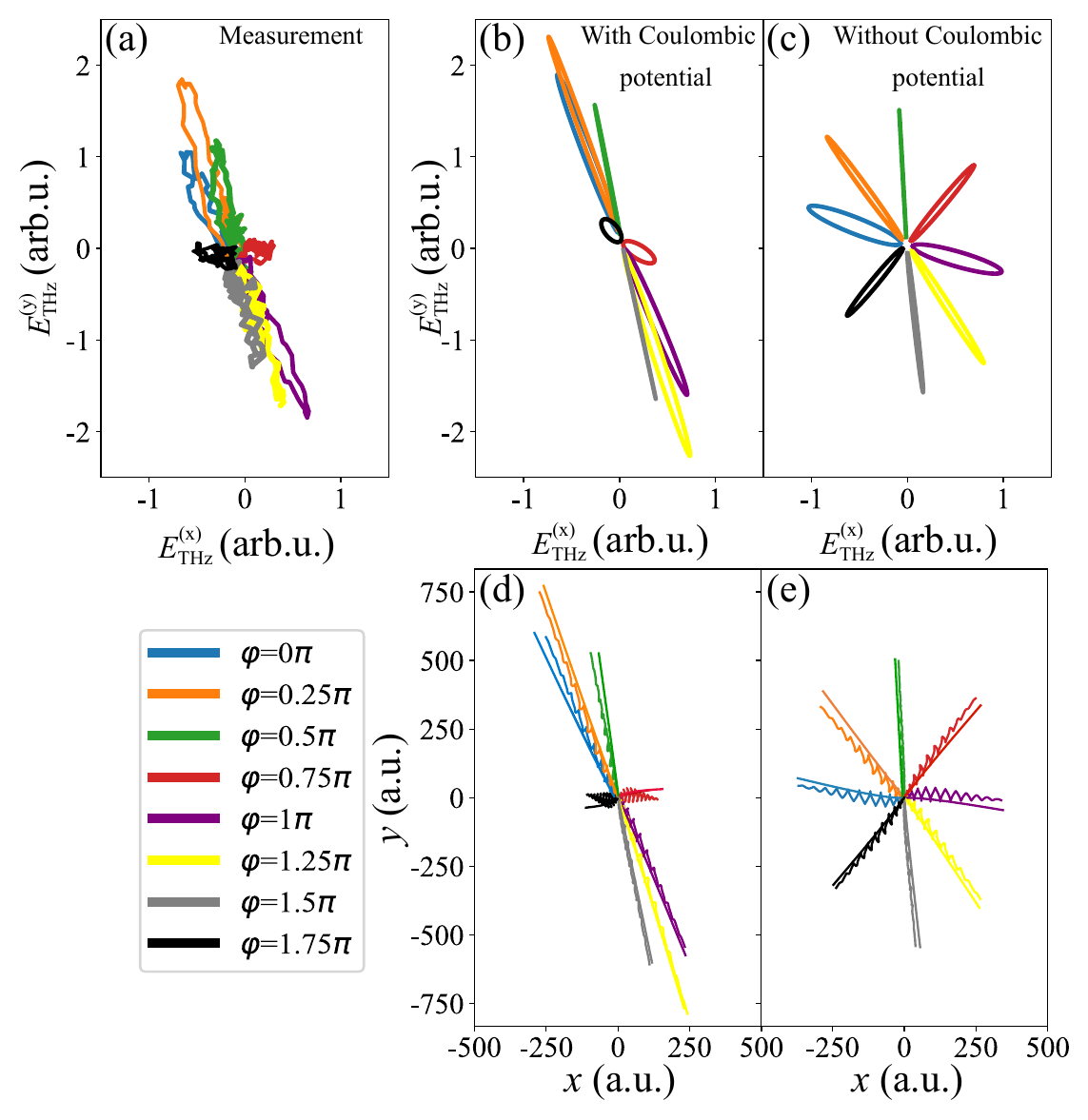}
\caption{The THz polarizations and electron trajectories as a function of phase delay $\tau$ under the linearly polarized $\omega$ field and circularly polarized 2$\omega$ field. (a) The experimental measurement of instantaneous vector trajectories of THz electric fields. (b) and (c) The predictions for THz polarizations by means of the CTMC simulations with and without Coulomb potential. (d) and (e) The CTMC-predicted electron trajectories $\boldsymbol{r}$ with and without Coulomb potential. The solid lines are the slowly-varying components of $\boldsymbol{r}$.}
\label{fig: Terahertz polarization}
\end{figure}

The instantaneous electric-field vector can be reconstructed from \begin{math}\boldsymbol{E}_{\text{THz}}\end{math}, experimentally and theoretically, shown in Fig. \ref{fig: Terahertz polarization}, where the property of the THz polarization can be feasibly obtained. Fig. \ref{fig: Terahertz polarization}(c) shows the CTMC simulation without Coulomb potential, which is equivalent to the photocurrent simulation \cite{Fan:23}, 
that the THz polarizations rotate with $\varphi$ \cite{2009Coherent1}. However, in our measurement (Fig. \ref{fig: Terahertz polarization}(a)), we found that the orientation of the THz polarization points to the fixed direction without dependence on $\varphi$. The photocurrent model and the CTMC without Coulomb effect are not consistent with the experimental result, whereas only the CTMC simulation considering Coulomb potential is able to reproduce the result, which indicates that the Coulomb effect cannot be neglected in THz generation process. 

To further illustrate the effect of the Coulomb potential, the electron trajectory analysis is implemented to correlate the trajectory behavior and the THz polarization, as shown in Figs. \ref{fig: Terahertz polarization}(d) and (e). Here, we show the equivalent trajectories at specific $\varphi$, which are obtained by averaging $10^{5}$ trajectories at different tunneling instants as $\sum\limits_\eta^{} {{W^{(\eta )} (t_0,p_\bot)} \boldsymbol{r}}^{(\eta )}$. The THz radiation can be correlated with the electron trajectory via the formula
\begin{equation}
    \boldsymbol{E}_\text{THz} \propto \lim_{\omega \rightarrow 0} \int_{-\infty}^\infty \boldsymbol{\Ddot{r}}(t) e^{-i \omega t}dt =\boldsymbol{\dot{r}} (\infty),
\end{equation}
which means that the orientation of the THz electric field is approximated to the direction of asymptotic velocity $\boldsymbol{\dot{r}} (\infty)$. Comparing the electron trajectories with and without Coulomb potential, it can be concluded that, the Coulomb scattering modifies $\boldsymbol{\dot{r}} (\infty)$, leading to the twist of the THz polarization.

Because the THz generation stems from the low-frequency behavior of the electron trajectory, it is speculated that the THz polarization can be analysed by the guiding center of electron trajectory \cite{Dubois2018}, which is obtained by removing the rapidly-varying laser-driven quiver motion from $\boldsymbol{r}$. The guiding center of electron trajectory, $\tilde{\boldsymbol{r}} (t)$, can be obtained as
\begin {equation} 
\tilde{\boldsymbol{r}} (t)=-\int_{t_0}^{t}d{t^{'}}\int_{t_0}^{t^{'}}d{t^{''}}\boldsymbol{E}_\text{THz}({t^{''}}). 
\end {equation}
As plotted as solid lines in Figs. \ref{fig: Terahertz polarization}(d) and (e), the asymptotic direction of $\tilde{\boldsymbol{r}}$ does approximately coincide the orientation of the THz polarization.

\begin{figure*}[h]
\centering
    \includegraphics[width=0.8\linewidth]{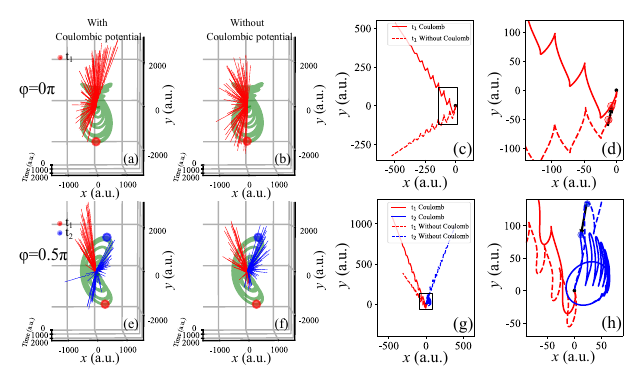}
\caption{Electronic trajectory distributions
 with/without core potential. (a) and (b) Electron trajectory distributions at $\varphi=0$ with/without Coulomb potential. The green line represents the electric field. (c) Electron trajectory with maximum ionization rate at $\varphi=0$. (d) Zoom-in  region near the core in (c). The black arrow is the direction of Coulomb force of parent core. (e) and (f) Electron trajectory distributions at $\varphi=0.5\pi$ with/without Coulomb potential. (g) Trajectory with maximum ionization rate at \begin{math} \varphi=0.5\pi \end{math}. (h) Zoom-in  region near the core in (g).
}
\label{fig: Electronic trajectory analysis}
\end{figure*} %

Fig. \ref{fig: Electronic trajectory analysis} shows the overall morphology of electron trajectory ensemble, which approximately reflects the temporal-spatial evolution of the photoelectron wavepacket. The upper row of Fig. \ref{fig: Electronic trajectory analysis} shows the trajectory analysis at $\varphi = 0$, where only one extremum appears within one cycle of the electric field. Without Coulomb potential (Fig. \ref{fig: Electronic trajectory analysis}(b)), the trajectory ensemble is distributed as a sector shape symmetric about $x$ axis. Fig. \ref{fig: Electronic trajectory analysis}(a) shows the effect of Coulomb scattering, which obviously twists the trajectory ensemble clockwise, and breaks the symmetry of the trajectory distribution. Figs. \ref{fig: Electronic trajectory analysis}(c) and (d) show the characteristic trajectories released at ionization instant $t_1$, when the trajectories have the maximum ionization weights. As shown in Fig. \ref{fig: Electronic trajectory analysis}(d), which plots the trajectory propagation within the beginning few cycle, the Coulomb force strongly decelerates the electron within the first half cycle and twists the direction of the electron motion. The Coulomb effect deflects $\boldsymbol{\dot{r}} (\infty)$ with $\sim 60^{\circ}$, leading to the rotation of the THz polarization.

The lower row of Fig. \ref{fig: Electronic trajectory analysis} shows the electron trajectories at $\varphi = 0.5\pi$, where two electric-field extrema within one cycle appear at $t_1$ and $t_2$. Without the Coulomb-potential impact (Fig. \ref{fig: Electronic trajectory analysis}(f)), there are two clusters of trajectories respectively emitting from the two extrema, which are symmetric about $y$ axis. If the Coulomb potential is taken into account (Fig. \ref{fig: Electronic trajectory analysis}(e)), whereas the left cluster of trajectories is rotated clockwise by Coulomb scattering, the right cluster experiences a recollision and is divided into two clusters.

Figs. \ref{fig: Electronic trajectory analysis}(g) and (h) show the high weight trajectories at ionization instant $t_1$ and $t_2$. The red trajectory released at $t_1$ undergoes a soft collision and clockwise rotation, which has similar behavior with the trajectory at $\varphi=0\pi$, whereas the blue trajectory released at $t_2$ is subjected to a recollision with parent core. Under attraction of Coulomb force, the electron oscillates back and forth closely around the core, and then rotates around the core to form a Rydberg state. Most of the electron trajectories emitted around $t_2$ with high ionization weights are recaptured by the core, which suppresses the THz radiation.

Comparing to the short-wavelength laser field, who drives the electron oscillating around the core, the long-wavelength field can pull the electron rapidly away from the core, which relieves the influence of Coulomb potential. 
In this case, since the Coulomb effect can be neglected, the THz radiation predicted by photocurrent model could be approximately equivalent to the CTMC without Coulomb interaction. In order to confirm the hypothesis, a 3900 nm 10-cycle linearly polarized field and a 1950 nm 20-cycle circularly polarized field are used in the theoretical simulation.

\begin{figure}[h]
\centering
    \includegraphics[width=0.6\linewidth]{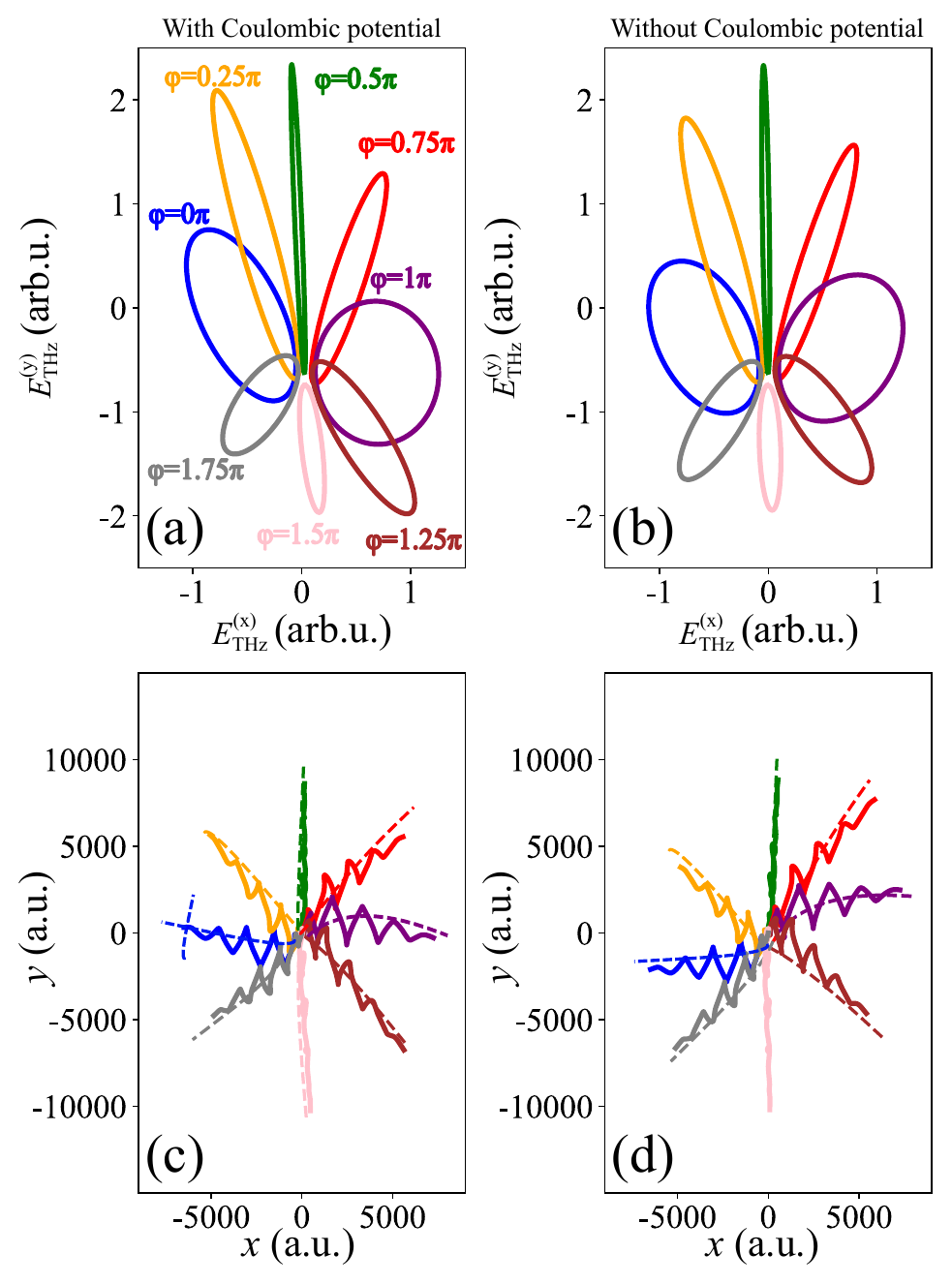}
\caption{THz polarization generated by 1950 nm circularly polarized laser field and 3900 nm linearly polarized laser field. (a) and (b) THz polarization calculated by CTMC with/without Coulomb potential. (c) and (d) Corresponding electron trajectories with/without Coulomb potential. The dashed line describes the guiding center of electron trajectory.
}
\label{fig: mesh7}
\end{figure} 

Fig. \ref{fig: mesh7} shows the simulation results of the instantaneous vector trajectories of THz electric fields and corresponding electron trajectories $\boldsymbol{r}$. The CTMC without Coulomb potential, shown in Fig. \ref{fig: mesh7}(b), is exactly equivalent to the photocurrent model, because they have the same analytic forms \cite{Fan:23}. As shown in Figs. \ref{fig: mesh7}(a) and (b), comparing to the linearly polarized THz radiation generated by the short-wavelength field, the orientation and ellipticity of the THz radiation generated under long-wavelength field can be controlled by varying the two-color phase delay $\varphi$, and the behavior of the THz radiation is not substantially modified by the Coulomb effect, which verifies our hypothesis.

The THz polarization can be indicated by the guiding center of the electron trajectory, as shown in Figs. \ref{fig: mesh7}(c) and (d). The guiding center of the trajectory has the transverse displacement perpendicular to the motion direction, as plotted in the purple dash line. The transverse displacement of the guiding-center trajectory is more pronounced in long-wavelength field than that in short-wavelength field (Figs. \ref{fig: Terahertz polarization}(d) and (e)). The transverse displacement creates the transverse acceleration perpendicular to the direction of $\boldsymbol{\dot{r}}(\infty)$, corresponding to the short axis of time trace of THz electric field. Hence, the large transverse displacement of guiding-center trajectory refers to the large ellipticity of THz electric field.

\section{Conclusion}
In conclusion, the Coulomb potential is found to be able to twist the THz polarization, which may benefit the application in the optical attoclock \cite{babushkin_2022}. The Coulomb-corrected CTMC method can reproduce the behavior of THz polarization generated experimentally by the linearly-polarized 800 nm and circularly-polarized 400 nm fields, which cannot be predicted by photocurrent method. Present trajectory analysis of Coulomb-corrected CTMC method shows that the elastic scattering of parent core modifies the orientation of THz electric field by twisting the asymptotic momentum of electron trajectory, and the core recaptures the electron, which impedes the THz radiation. The CTMC simulation predicts that the two-color mid-infrared fields can effectively accelerate the electron rapidly away from the parent core to elude the influence of Coulomb potential. And because the slowly-varying envelope of two-color long-wavelength fields produces the lateral momentum perpendicular to asymptotic momentum of guiding-center electron trajectory, the mid-infrared fields are able to generate the circularly-polarized THz radiation, whose orientation and ellipticity can be manipulated via two-color phase delay. 

\section*{Funding}
This work was supported by the National Key Research and
Development Program of China (No. 2022YFA1604302) and National Natural Science
Foundation of China (NSFC) (Nos. 12174284 and 11827806).

\section*{Disclosures}
The authors declare no conflicts of interest.

\bibliographystyle{unsrt}
\bibliography{main}

\newpage
\section*{EXPERIMENTAL SETUP} 
.\begin{figure*}
\centering
    \includegraphics[width=\linewidth]{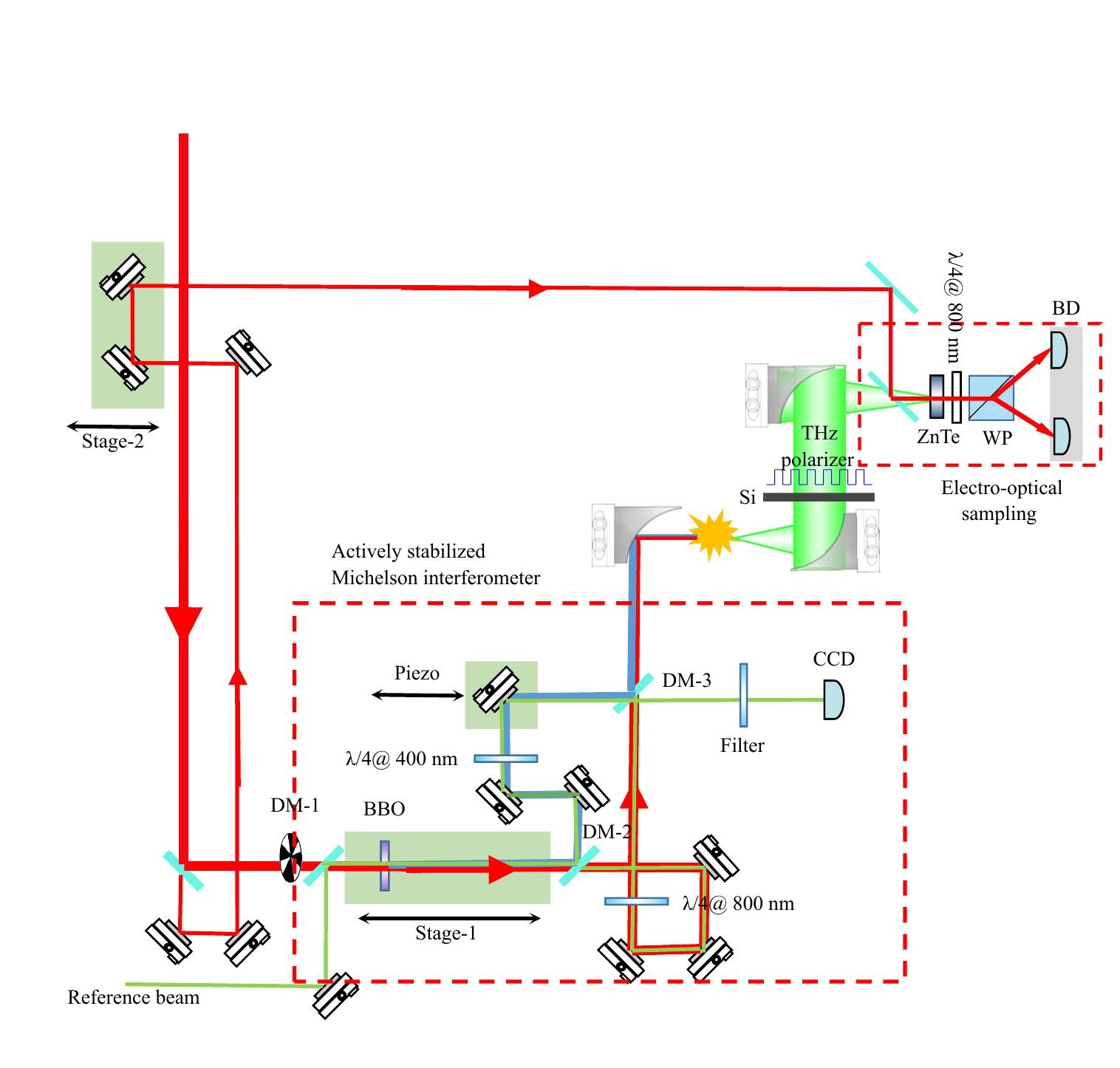}
    \caption{Schematic of the experimental setup. The focused pump beam passes a \begin{math}\beta-BBO\end{math} crystal, creating plasma at the focus. THz pulse is detected by a standard electric-optic sampling setup.}
\label{fig: Experimental setup diagram}
\end{figure*}

The experimental apparatus is illustrated in Fig. \ref{fig: Experimental setup diagram}. A Ti:sapphire laser delivers an $x$-polarized 35 fs, 1.2 mJ femtosecond
pulse centered at 800 nm. The 800 nm fundamental-frequency ($\omega$) beam passes
through a 200 µm type-I $\beta$-barium borate (BBO) crystal, and a $y$-polarized 400 nm second-harmonic ($2\omega$) beam is generated (conversion efficiency $\sim$ $30\%$). The co-propagating two-color laser pulses are separated by a dichroic mirror (DM-2) into the two arms of a Michelson interferometry. The polarization of the $\omega$-$2\omega$ beams can be independently controlled by quarter-waveplates ($\lambda$/4$@$400, $\lambda$/4$@$800). During the measurement, the $\omega$ electric field is longitudinally aligned along the $x$-axis, and the 2$\omega$ electric field is circularly polarized.

The phase stabilization between the $\omega$-$2\omega$ beams is controlled by implementing an actively stabilized Michelson interferometer. The stabilization of the relative phase of the two arms in the Michelson interferometer is performed by introducing another continuous green laser (532 nm) as a reference beam, who co-propagates along with an $\omega$-$2\omega$ beam. 

The reference beam and femtosecond beams are combined by DM-1, and the reference beam is separated and recombined by DM-2 and DM-3. The CCD detects the interference fringes of the reference beam, and a piezo provides feedback to actively lock the phase between the two paths of the Michelson interferometer to stabilize the interference fringes. During data acquisition, the relative phase fluctuation of the stabilized system is $<0.02\pi$. The phase delay $\varphi$ can be adjusted with sub-femtosecond accuracy by changing the distance between the BBO on Stage-1 and air plasma, due to the different refractive indices of the $\omega$-$2\omega$ fields in air.

After the DM-3, the $\omega$-$2\omega$ fields collinearly propagates, and are focused by a parabolic mirror with focus length of 100 mm, which induces the plasma. The THz beam emitted from the plasma is aligned by two parabolic mirrors and focused on a 1 mm thickness ZnTe crystal to implement the electro-optical sampling (EOS) techniques. In the EOS, an 800 µm thickness silicon wafer blocks the $\omega$ and $2\omega$ beams. The ZnTe crystal fixed at the special orientation, which has the same responses for $x$- and $y$-polarized components, accompanies a THz wire grid polarizer to resolve both orthogonal polarizations, $E_{\mathrm{THz}}^{\mathrm{(x)}}$ and $E_{\mathrm{THz}}^{\mathrm{(y)}}$. The time trace of the THz instantaneous electric field can be reconstructed with $E_{\mathrm{THz}}^{\mathrm{(x)}}$ and $E_{\mathrm{THz}}^{\mathrm{(y)}}$ \cite{zhang_portraying_2009}.

\end{document}